\documentclass[11pt,a4paper]{article}
\usepackage{natbib}
\usepackage{amsthm}
\usepackage{enumerate}
\newtheorem{theorem}{Theorem}
\usepackage{geometry}
\usepackage{color}
\usepackage{hyperref}

\usepackage{rotating} \usepackage{graphicx} \usepackage{subcaption}

\usepackage{float} 
\usepackage{dsfont}
\usepackage{amsmath, amssymb} 
\usepackage{mathrsfs}
\usepackage{subcaption}
\usepackage[normalem]{ulem}

\usepackage{tikz}
\usepackage{tikz-3dplot}
\newcommand{\be}{\begin{equation*}\begin{aligned} }
\newcommand{\ee}{\end{aligned}\end{equation*} }

\newcommand{\bel}{\begin{equation}\begin{aligned} }
\newcommand{\eel}{\end{aligned}\end{equation} }

 \DeclareMathOperator{\No}{No}

 \DeclareMathOperator{\diag}{diag}

\newcommand{\T}{^{\rm T}}

\begin{document}
\title{Bayesian Multi-scale Modeling of Factor Matrix\\without using Partition Tree}

\author{Maoran Xu\thanks{Department of Statistics, University of Florida, Gainesville, FL, email: maoranxu@ufl.edu}
\quad \quad Leo L. Duan\thanks{Department of Statistics, University of Florida, Gainesville, FL, email: li.duan@ufl.edu}
}

\maketitle
\begin{abstract}
The multi-scale factor models are particularly appealing for analyzing matrix- or tensor-valued data, due to their adaptiveness to local geometry and intuitive interpretation. However, the reliance on the binary tree for recursive partitioning creates high complexity in the parameter space, making it extremely challenging to quantify its uncertainty. In this article, we discover an alternative way to generate multi-scale matrix using simple matrix operation: starting from a random matrix with each column having two unique values, its Cholesky whitening transform obeys a recursive partitioning structure. This allows us to consider a generative distribution with large prior support on common multi-scale factor models, and efficient posterior computation via Hamiltonian Monte Carlo. We demonstrate its potential in a multi-scale factor model to find broader regions of interest for human brain connectivity.

\end{abstract}
{\bf keywords: }Consistent Partitioning, Multi-scale Stiefel Manifold, Multi-scale Tensor Factorization.

\section{Introduction}
Factor models play a dominant role in multivariate statistics, as they reveal the low-dimensional structure hidden in the matrix- and tensor-valued data. For example, in the matrix factorization model, the noisy observed matrix $A\in \mathbb{R}^{n_1\times n_2}$ can be decomposed into a low-rank-plus-noise structure $A = UDV\T + E$, with $U=(\vec u_{1} \ldots \vec u_{k})\in \mathbb{R}^{n_1\times k}$, $V=(\vec v_1 \ldots \vec v_{k})\in \mathbb{R}^{n_2\times k}$, $D =\diag\{d_i\}_{i=1}^k$ and $E\in \mathbb R^{n_1\times n_2}$ the idiosyncratic error such as Gaussian noise. Similar low-rank decompositions are extended to modeling higher-order tensor data $A\in\mathbb{R}^{n_1\times n_2\times\cdots \times n_p}$. For reviews and various extensions, see \cite{fosdick2014separable,guhaniyogi2017bayesian}, among others.

When the dimension $n$ is large (omitting subscript now for simple notation), a multi-scale parameterization becomes appealing, as an effective way to further reduce the number of parameters while improving model interpretation. For example, in the above matrix factorization model, we can limit the number of unique values in each vector $\vec u_j$ and $\vec v_j$, creating block structure in each component matrix $\vec u_j \otimes \vec v_j$; by allowing more unique values for higher level $j$, we create a multi-resolution decomposition of the raw data. This idea has nice applications on real world data, such as image segmentation \citep{choi2001multiscale}.

One key issue, however, is to decide which elements should share the same value. Most of the existing approaches rely on a binary partition tree. Specifically, the tree recursively divides the row index set $\{1,\ldots,n\}$ of a factor matrix into at most $2^j$ disjoint subsets at level $j$; one can then parameterize $\vec u_{j}=(u_{i,j})_i$, by constraining $u_{i,j}=u_{i',j}$ if $i$ and $i'$ are in the same index subset. There is a sizable optimization literature on obtaining point estimate on the tree, including approximation via greedy and non-greedy algorithms \citep{breiman2017classification,norouzi2015efficient}. On the other hand, it is extremely challenging to quantify the uncertainty --- in particular, if we equip the partitioning with a prior distribution, then (i) how much can the partitioning vary within the high posterior probability region? (ii) does the posterior distribution governing the partitioning concentrate as the data sample size increases?

Some early works \citep{chipman1998bayesian,wu2007bayesian} have attempted to address (i) under a regression tree context, using a depth-penalizing prior and Markov chain Monte Carlo algorithm to explore local changes to the tree. Unfortunately, those discrete proposals were found inefficient for exploring the partition space, with the Markov chain quickly stuck near a local optimum. For (ii), there have been posterior consistency results established in terms of  density estimation using Bayesian tree models \citep{wong2010optional,ma2017adaptive}, yet to our best knowledge, it is not clear whether the partitioning itself can be consistently estimated, which is crucial for interpretation in the multi-scale factor analysis.

We address these issues by completely removing the reliance on the tree, hence substantially reduce the model complexity for Bayesian multi-scale methods. Our method was inspired by the recently proposed polar transform when re-parameterizing an orthogonal matrix $Q_X$ using an unconstrained $X\in \mathbb{R}^{n\times k}$ (Jauch et al. 2019 arXiv preprint 1906.07684); nevertheless, our unique contribution is in the discovery that the Cholesky whitening transform of a {\em structured} $X$ can lead to recursive partitioning structure in each column $Q_X$. This allows us to consider a generative model on the multi-scale orthogonal space, which enjoys efficient posterior estimation via Hamiltonian Monte Carlo, and prior support sufficient for establishing posterior consistency for common multi-scale factor models.

\section{Bayesian Modeling of Multi-scale Orthogonal Matrix}

\subsection{Transforming to Multi-scale via Cholesky Whitening}

In order to formalize a multi-scale matrix $Q=(\vec q_1\ldots \vec q_k)\in \mathbb{R}^{n\times k}$, we first introduce some set notations. Denote $(C_{(j)1},C_{(j)2})$ a bi-partition of $[n]=\{1,\ldots,n\}$ at level $j$, such that $C_{(j)1} \cup C_{(j)2}=[n]$ and $C_{(j)1} \cap C_{(j)2}=\varnothing$. A $k$-level recursive binary partition starts from $V_{(0)1}=[n]$, and then recursively takes an existing $V_{(j-1)l}$ and divides into $V_{(j)l'}= V_{(j-1)l} \cap C_{(j)1} $ and $V_{(j)l''}= V_{(j-1)l} \cap C_{(j)2}$, for $j=1,\ldots,k$ and for $l\in\{1,2\}^{j-1}$. Clearly, $\cup_{l\in\{1,2\}^j} V_{(j)l}=[n]$ and $V_{(j)l}\cap V_{(j)l'}=\varnothing$ for $l\neq l'$.

It is not hard to see that each resulted partition can be succinctly re-presented as $V_{(j)l}=C_{(1)h_1} \cap C_{(2)h_2} \cap \cdots \cap C_{(j)h_j}$ for $l=(h_1,h_2,\ldots,h_j )\in \{1,2\}^j$. Since the intersection is allowed to be empty, we can have fewer than $2^j$ partitions at level $j$. As the last step, we parameterize $q_{i,j}=q_{i',j}$ if $i\in V_{(j)l}$ and $i'\in V_{(j)l}$. In the tree-based methods, one would treat $h_j$ as the left or right split of a node.

We consider an alternative that achieves the same recursive partition structure. Consider a matrix $X\in \mathbb{R}^{n\times k}$, with each column having only two unique values: 
\be
x_{i,j}=a_j \text{ if }i\in C_{(j)1},\\
x_{i,j}=b_j\text{ if }i\in C_{(j)2},
\ee
 where $a_j\neq b_j$. Assuming $n\ge k$ and $X$ has full rank, we apply the Cholesky whitening transform:
\be
Q = X \{\mathcal{L}(X\T X)\}^{-\rm T},
\ee
where $\mathcal{L}:\mathbb{S}^{k}_{++}\to \mathbb{R}^{k\times(k+1)/2}$ is the Cholesky decomposition that takes positive definite $S$ and produces lower-triangular matrix $L$ such that the diagonal values are positive and $S=LL\T$; $()^{-\rm T}$ denotes the inverse transpose.  It is not hard to see that $Q$ is in a Stiefel manifold $\mathcal V^{k, n}=\{Q\in\mathbb R^{n\times k}: Q^{\rm T}Q=I_k\}$. More importantly, this simple matrix operation produces a recursive partition structure.

\begin{theorem}[Multi-scale partition on the Stiefel manifold]
For any  $j=1,\ldots,k$, $l\in\{1,2\}^j$, $i\in [n], i'\in[n]$, the Cholesky whitening transform has $q_{i,j}=q_{i',j}$ if $i\in V_{(j)l}$ and $i'\in V_{(j)l}$. And there are at most $2^j$ unique values in $\vec q_j$.
\end{theorem}

\begin{figure}[h]
\begin{center}
 \includegraphics[width=3in]{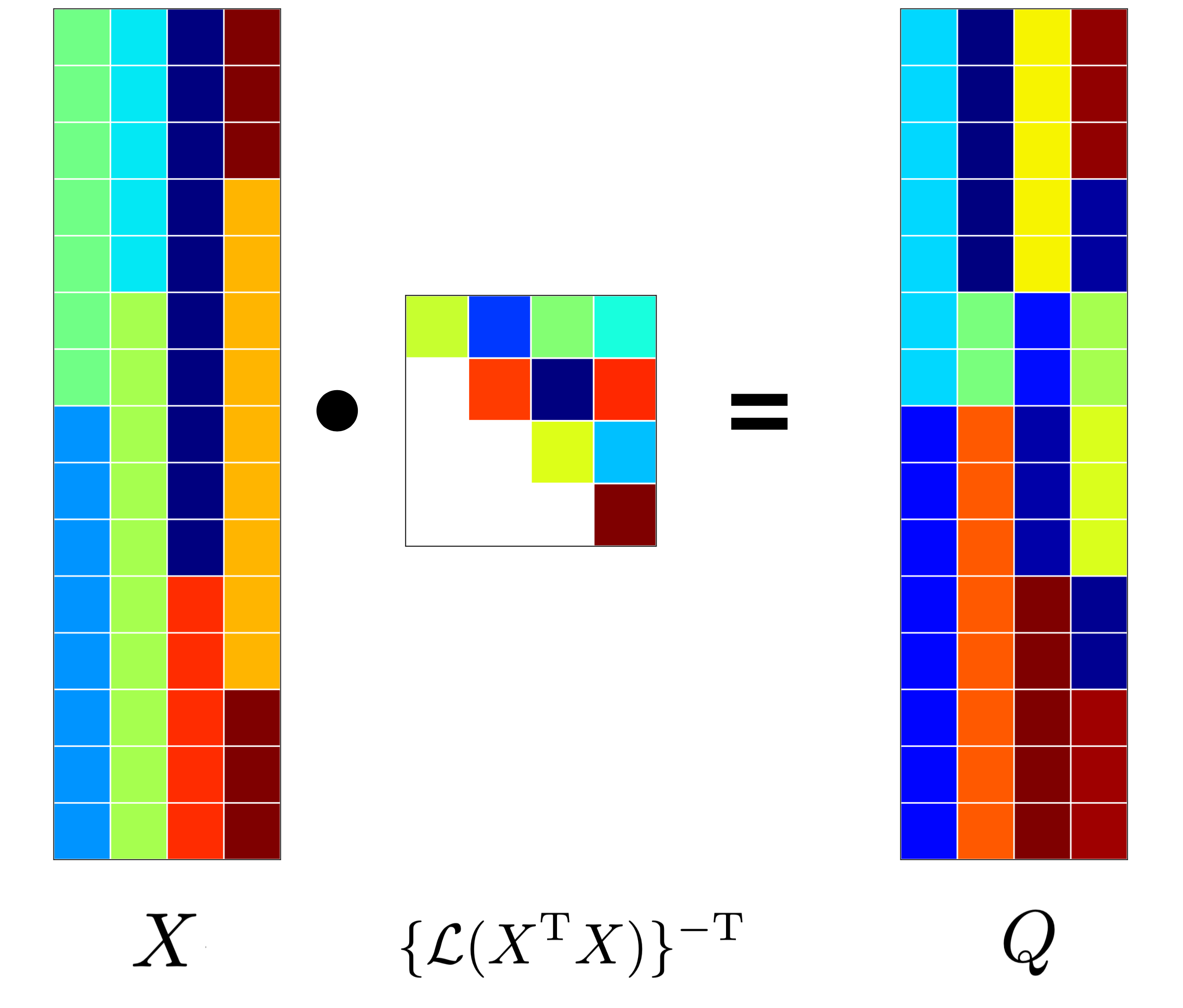}
\end{center}
   \caption{Illustration of how the Cholesky whitening transform converts the matrix $X$ into a matrix $Q$, that has at most $2^j$ unique values in the $j$th column and obeys a multi-scale structure.}
\end{figure}

\subsection{Multi-scale Orthogonal Prior}
To exploit the above result, we now propose a generative distribution for $Q$ to be used as prior in Bayesian models:
\bel\label{eq:chol_model}
& a_j\sim \No(0,1), \quad b_j\sim \No(0,1), \quad p_j\sim \text{Beta}(1,1) \text{ for } j =1,\ldots,k,
\\
& \pi_{X\mid (\vec a,\vec b,\vec p)}(X) =Z^{-1}_{(\vec a,\vec b,\vec p)} \bigg [ \prod_{j=1}^k \prod_{i=1}^n \big \{ p_j \delta_{a_j}(x_{i,j}) + (1-p_j) \delta_{b_j}(x_{i,j}) \big \} \bigg] \mathbb{I}\{\text{rank}(X)=k \},\\
& Q = X \{\mathcal{L}(X\T X)\}^{-\rm T},
\eel
where vectors $\vec a=(a_j)_j$, $\vec b=(b_j)_j$ and $\vec p=(p_j)_j\in \mathbb{R}^{k}$; $\delta_{c}(.)$ denotes the point mass at point $c\in \mathbb{R}$; $ \mathbb{I}_E$ is the indicator taking value $1$ if condition $E$ holds, otherwise taking $0$; $Z$ is the normalizing constant that depends on $(\vec a,\vec b,\vec p)$. Since $Q$ is invariant to the re-scaling of $X$, we fix the variances of $a_j$ and $b_j$ to $1$, and use non-informative prior for $p_j$. We refer to the induced distribution of $Q$ as the multi-scale orthogonal prior.

To derive its prior density, first note that although the conditional $X\mid (\vec a,\vec b,\vec p)$ follows a discrete distribution, the marginal $X$ is continuous and has density $\pi_X(X)=  \int \pi_{X\mid (\vec a,\vec b,\vec p)}(X)  \pi_{(\vec a,\vec b,\vec p)}( \vec a,\vec b,\vec p)\text{\,d} (\vec a,\vec b,\vec p)$. Applying variable substitution  $X\to(Q, M)$, with $M=X\T X$, we have
\be
\pi_X(X) = \pi_{Q,M}(Q,M)\frac{\Gamma_k(n/2)}{\pi^{nk/2}}(\det M)^{{-(n-k-1)}/{2}},
\ee
where the last part corresponds to the Jacobian determinant of the transformation (Lemma 1.5.2, \cite{chikuse2012statistics}). Since our interest is on $Q$, we write the marginal as
\bel\label{eq:marginal_prior}
 \pi_Q(Q) = \int_{\mathbb{S}^{k}_{++}}\! \pi_X \{Q\mathcal {L}\T (M)\}
 \frac{\pi^{nk/2}}{\Gamma_k(n/2)}
 (\det M)^{{(n-k-1)}/{2}}
 \text{\,d}M.
\eel
Despite the lack of closed-form marginal, we can draw $(\vec a, \vec b, \vec p, X)$ together, and then obtain $Q$ as a deterministic transform of $X$.

Using $\pi_Q(Q)$ as a prior, $L(A;Q,\theta)$ as the likelihood for data $A$, $\pi_\theta(\theta)$ as the prior for other parameter $\theta$, we have the posterior
\[
\pi(Q,\theta\mid A) \propto L(A;Q,\theta)\pi_Q(Q) \pi_\theta(\theta).
\]
We will describe the sampling algorithm later, including how to deal with the intractable constant $Z$. For now, we establish the statistical property of the proposed prior.

\subsection{Prior Support and Posterior Consistency}

A primary consideration is if this prior has enough support to allow a consistent estimation of the partitioning. We now show this is achievable in the context of estimating the multi-scale factor matrix $Q$. Specifically, with the data as a sequence $A^{(n)}=\{A_1,A_2,\ldots, A_n\}$, with $A_s\sim  f_{Q_0}$ independently for $s = 1,\ldots ,n$ and for some fixed true value $Q_0$, slightly abusing notation, we denote $L(A^{(n)};Q)  =\prod_{i=1}^n f_{Q}(A_i)$ as the joint likelihood, marginalized over $\theta$. As $n\to \infty$, we hope the posterior $\pi(Q\mid A^{(n)})$ concentrates in a small neighborhood of $Q_0$. Without imposing restrictive assumptions on $f$, we focus on the weak neighborhood
\be
B_\epsilon(f_{Q_0}) = \bigg\{f_Q:\bigg |\int\! g f_Q \mu({\textup d} Q) - \int \!g f_{Q_0} \mu( {\textup d} Q) \bigg| \le \epsilon, ~ \forall  g\in C_b(\mathcal
V^{k, n}) \bigg\},
\ee
with $C_b$ the class of continuous and bounded functions, and $\mu(.)$ the Haar measure on $\mathcal V^{k, n}$. A sufficient condition for the posterior consistency
\be
\Pi\left\{B_\epsilon(f_{Q_0})  \mid A^{(n)}\right\} \to 1,
\ee
almost surely with respect to $P_{f_{Q_0}}^{n}\text{\,as } n\to \infty$ is that $f_{Q_0}$ is in the Kullback-Leibler support of the prior \citep{ghosh2003bayesian}. We list the mild conditions for $f_Q$ in the following theorem.
\begin{theorem}
If $f_Q$ is uniformly continuous in $Q$, and  $\int f_{Q_0}|\log(\sup_{Q\in\mathcal{V}^{k,n}} f_Q)-\log(\inf_{Q\in\mathcal{V}^{k,n}} f_Q)|\lambda(\textup d A)<\infty$, with $\lambda$ the appropriate measure for $A$, then  $f_{Q_0}$ is in the Kullback-Leibler support of the prior $\Pi$ as in \eqref{eq:marginal_prior}. That is, for any $\epsilon>0,$
$$\Pi\left[ \{ f_Q:\int\! f_{Q_0}\log\frac{f_{Q_0}}{f_Q}  \lambda( {\textup d} A) <\epsilon \} \right]  >0.$$
\end{theorem}

As our prior makes it easy to adopt multi-scale extension to common factor models, we found a large class of them that satisfy the conditions above. Examples include the model-based Tucker decomposition \citep{hoff2016equivariant} and the dependence latent space model for multiple networks \citep{salter2017latent}. We will illustrate one example in Section 4.

\section{Posterior Sampling}
We use Markov chain Monte Carlo to sample from the posterior distribution. 
In order to deal with the intractable $Z$, we re-parameterize $X=W\diag\{a_j\}+(J-W)\diag\{b_j\}$ using a binary matrix $W\in\{0,1\}^{n\times k}$, with $J$ an $n\times k$ matrix filled by ones. 

This allows us to write the joint posterior of all parameters as
\be
\pi &(\theta,X, \vec a, \vec b, \vec p \mid A) = c \frac{\exp\{ -U (\theta,W; \vec a, \vec b, \vec p)\}}{Z(\vec a, \vec b, \vec p)}  \mathbb{I}\{\text{rank}(X)=k \},\\
U&(\theta,W; \vec a, \vec b, \vec p) = 
-\log L(A;Q_X,\theta)
-\log \pi_\theta(\theta)
-\log g(W;\vec p) \\
&\qquad- {\frac{(n-k-1)}{2}}\log\det M
+ \sum_j (
a_j^2/2+ b_j^2/2),\\
 g& (W;\vec p)= \prod_j \prod_i
p_j^{ w_{i,j}}
(1-p_j)^{(1- w_{i,j})},
\ee
where $c$ is a constant that does not depend on the parameters. For simplicity, we assume $L(A;Q,\theta)$ and $\pi_\theta$ do not involve intractable normalizing constant.

We use two steps in each sampling iteration:

	\quad (i) Update $(W,\theta) \mid (\vec a, \vec b, \vec p)$ using Hamiltonian Monte Carlo;

	\quad (ii) Update $ (\vec a, \vec b, \vec p) \mid (W,\theta)$ using the exchange algorithm Monte Carlo \citep{murrayexchange}.

In step (i), since $W$ could be in high dimension, we use the Hamiltonian Monte Carlo with continuous approximation to more efficiently explore the binary space. Following \cite{maddison2016concrete}, we approximate $w_{i,j}$ by $\tilde w_{i,j} = 1/\{1+\exp(-\beta_{i,j}/\tau)\}$, with $\beta_{i,j}\in\mathbb{R}$ and $\tau>0$ and close to $0$; this makes $\tilde w_{i,j}$ numerically very close to $0$ or $1$ with an overwhelming probability, yet differentiable with respect to $\beta_{i,j}$, hence suitable for simulating Hamiltonian dynamics.  An alternative implementation could be the Discontinuous Hamiltonian Monte Carlo \citep{nishimura2017discontinuous}. For step (ii), we utilize the exchange algorithm to effectively cancel out $Z$ in the Metropolis-Hastings. The sampling algorithm runs very efficiently, showing rapid mixing in the Markov chain. For conciseness, we defer all the algorithmic details to the appendix.

\section{Example: Finding Multi-scale Factors from Brain Network Data}

We demonstrate the potential of our proposed framework in a network data analysis. The data are pairwise connectivities of the human brain networks collected from a working memory study, and represented by symmetric binary matrices $A_s\in \{0,1\}^{n\times n}$, with $n=128$ over  subject $s=1,\ldots,20$. In neuroscience, a primary interest is to identify the broad regions of the brain that contain most of the connectivities, hence the multi-scale model can be particularly useful.

\begin{figure}[H]
\begin{subfigure}[b]{.3\textwidth}
  \centering
 \includegraphics[width=1.3in, height=1.3in]{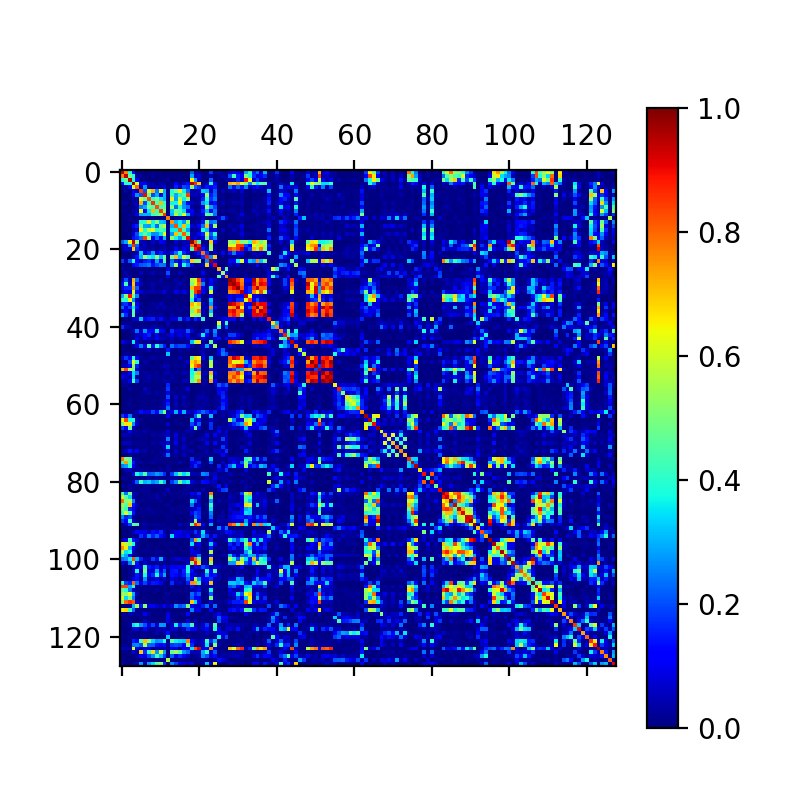}
 \caption{The average connectivity matrix $\bar A$ from $20$ subjects, collected during a working memory study.}
\end{subfigure}
\begin{subfigure}[b]{.6\textwidth}
  \centering
 \includegraphics[width=1\textwidth, height=1.3in]{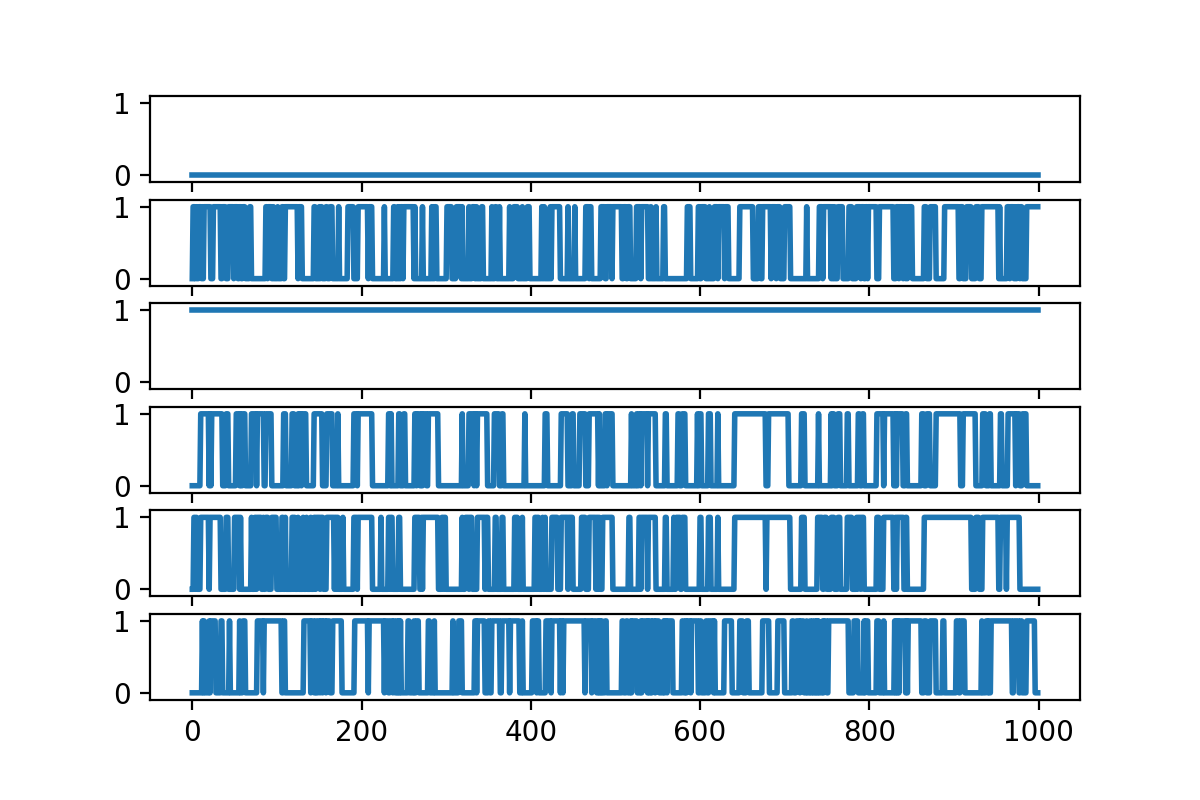}
    \caption{Rapid mixing of Markov chain for estimating the partition assignments: $6$ nodes are shown in terms of $\mathbb{I}(x_{i,3}=a_3)$, among which node $2,4,5,6 $ have $\text{pr}(x_{i,3}=a_3)$ away from $0$ or $1$.}
\end{subfigure}
\begin{subfigure}[b]{.3\textwidth}
  \centering
 \includegraphics[width=1.3in, height=1.3in]{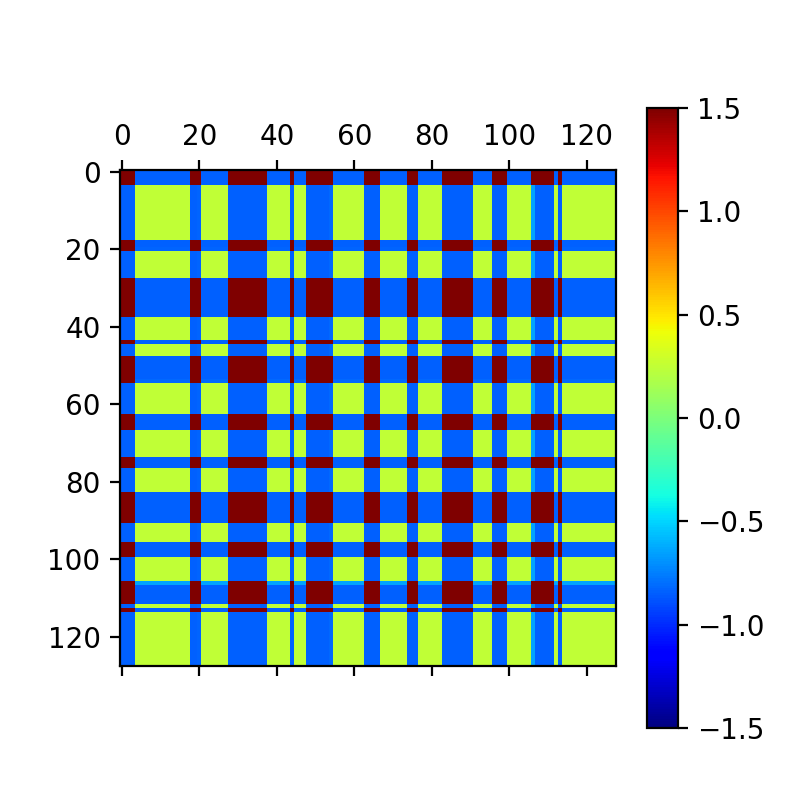}
  \caption{Factor 1: $\bar d_1\vec q_1 \vec q_1\T$.}
\end{subfigure}
\begin{subfigure}[b]{.3\textwidth}
  \centering
 \includegraphics[width=1.3in, height=1.3in]{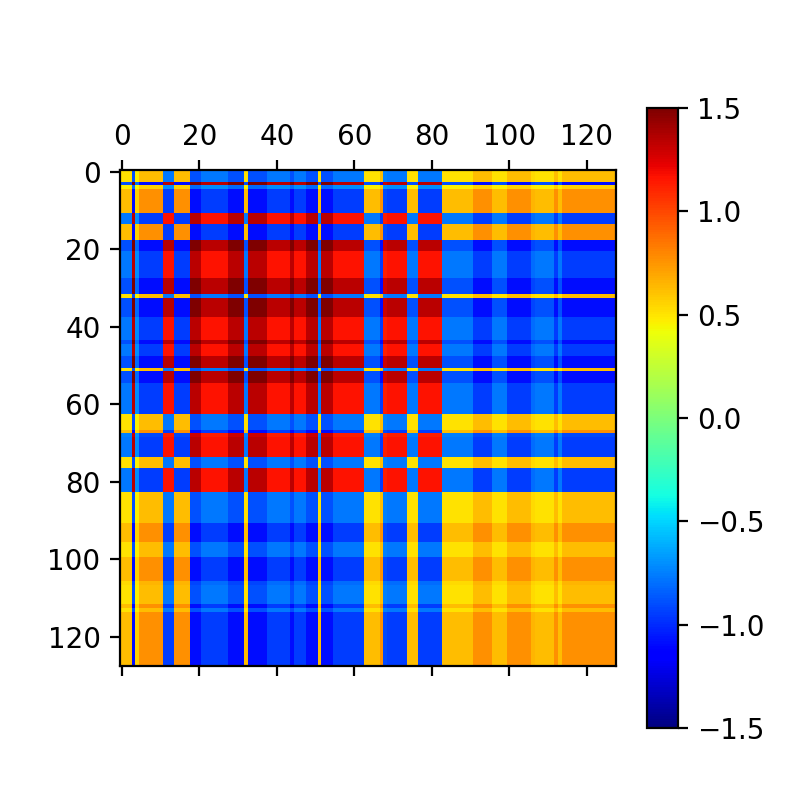}
   \caption{Factor 2: $\bar d_2\vec q_2 \vec q_2\T$.}
\end{subfigure}
\begin{subfigure}[b]{.3\textwidth}
  \centering
 \includegraphics[width=1.3in, height=1.3in]{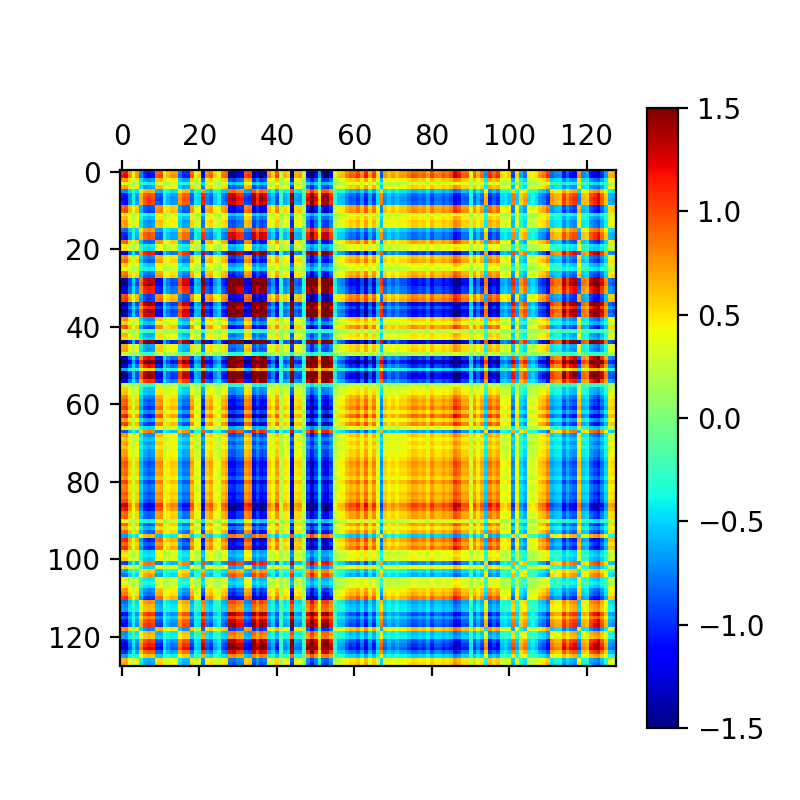}
    \caption{Factor 3: $\bar d_3\vec q_3 \vec q_3\T$.}
\end{subfigure}
 \caption{Illustration of estimating multi-scale latent factors from a group of connectivity matrices. \label{fig:network}}
 \end{figure}

We consider the following logistic Bernoulli latent factor model:
\be
& A_{s; i,j} \sim \text{Bernoulli} \big\{ \frac{1}{1+\exp(-\psi_{s;i,j})} \big\},\\
& \psi_s = Q D_s Q\T + z_s,
\ee
where $Q\in \mathcal V^{k,n}$ is shared among all subjects, and assigned a multi-scale orthogonal prior;  $D_s=\diag\{d_{s;j}\}$ is the individual loading on the factors and we assign inverse-Gamma$(0.1,0.1)$ prior for each $d_{s;j}$; $z_s$ is a scalar addressing the random fluctuation of the sparsity levels among matrices $A_s$. Based on the exploratory data analysis, we found there are about $30$ dominating eigenvalues in the spectrum of $\log\{\bar A/(1-\bar A)\}$, with $\bar A$ the average of $A_s$'s as a rough estimate for connectivity probability, therefore we choose $k=30$ and present the first few factors. Figure~\ref{fig:network} shows how the connectivity probability matrix (panel a) can be decomposed into several major factor matrices at increasing resolutions (panel c-e, on log-odds scale) --- among which, the first factor corresponds to the connectivities within the forehead area; the second corresponds to the connectivities within the left and right hemispheres. The traceplots (panel b) show that the Markov chain mixes rapidly, providing uncertainty estimate on the partitioning. 
 
\section{Discussion}
Several extensions may be pursued based on our proposed framework: (i) the recursive bi-partition can be easily modified into a recursive $d$-partition (with $d>2$), creating a potentially more efficient multi-scale model with fewer scales; (ii) the Cholesky whitening transformed matrix can be adopted in the non-parametric framework, serving as a latent variable to induce multi-scale measures.

\appendix
\section*{Appendix}

%\numberwithin{equation}{section}
\subsection*{Proof of Theorem 1}
\begin{proof}
Let $L^*:=\{\mathcal{L}(X\T X)\}^{-\rm T}$, which is upper triangular. Therefore
$$q_{i,j}=\sum_{k=1}^jx_{i,k}L^*_{k,j}.$$
Since for each $k$, there are at most two unique values in $\vec x_k=(x_{i,k})_i$, there are at most $2^j$ unique values in $\vec q_j = (q_{i,j})_i$. When $i\in V_{(j)l}$ and $i'\in V_{(j)l}$, we have $x_{i,k}=x_{i',k}$, hence $q_{i,j}=q_{i',j}$.
\end{proof}

\subsection*{Proof of Theorem 2}
\begin{proof}

Let $D_\delta = \{Q: ||Q-Q_0||_F<\delta\}$. For any given $A$ and $\eta >0$, by uniform continuity of $f$, there exists a $\delta>0~$such that$~|f_Q(A)-f_{Q_0}(A)|<\eta$ whenever $Q\in D_\delta.$ 

Equivalently, $\log\frac{f_{Q_0}}{f_{Q}}\to 0$ as $Q\to Q_0$. Since $f_{Q_0}\log\frac{f_{Q_0}}{f_{Q}}$ is dominated by $ f_{Q_0}|\log(\sup_{Q\in D} f_Q)-\log(\inf_{Q\in D} f_Q)|$ (which is integrable by assumption), applying the dominated convergence theorem, it follows that
$$\int\! f_{Q_0}\log\frac{f_{Q_0}}{f_{Q}}\text{\,d}\lambda\to 0\text{~as~} Q\to Q_0.$$
That is, for any $\epsilon>0$, there exists a $\delta>0$ such that 
 \bel
\Pi \left[\left\{Q:\int\! f_{Q_0}\log\frac{f_{Q_0}}{f_{Q}}\text{\,d}\lambda<\epsilon\right\}\right]\ge \Pi(D_\delta).
\label{equ: 1}
\eel

Recall that $Q$ is transformed from $X$ by Cholesky whitening transform, where $X$ is assigned prior \eqref{eq:chol_model}. It is easy to show that for full rank $X$ and $X_0$, $Q\to Q_0$ as $X\to X_0$. Therefore 
\begin{equation}
	\Pi(D_\delta)\ge\pi_X\left\{\mathcal N_\gamma(X_0)\right\},
	\label{equ: 2}
\end{equation}
 where $\mathcal N_\gamma(X_0)$ is the F-norm neighborhood of radius $\gamma$ centered at $X_0$.

Under the re-parameterization $X(W,\vec a,\vec b) = W\text{diag}\{\vec a\}+(1-W)\text{diag}\{\vec b\}$, where $W_{ij}\sim \text{Bernoulli}(p_j)$, we rewrite $\pi_{X}(X)$ into the equivalent  joint distribution of $W,\vec a,\vec b$ as
 $$\pi(W,\vec a,\vec b) =\frac{1}{(2\pi)^{k}} \exp[ tr(W\log P)+tr\left\{(1-W)\log(J-P)\right\}]\exp\left\{-\frac{||\vec a||_2^2+||\vec b||_2^2}{2}\right\},$$
where $P\in \mathbb R^{k\times n}$ with row $j$ filled by $p_j\vec{1}^T$.
Then for fixed $W_0$, there exists a $\gamma^*$-radius neighborhood of $(\vec a_0, \vec b_0)$ such that $X(W_0,\vec a,\vec b)\in \mathcal N_\gamma(X_0)$ for $(\vec a,\vec b)\in \mathcal N_{\gamma^*}\{(\vec a_0,\vec b_0)\}.$ Then
\begin{equation}
\begin{aligned}
&\int_{N_\gamma(X_0)} \pi_X(X)dX\ge\int_{\mathcal N_{\gamma^*}\{(\vec a_0,\vec b_0)\}} \!\!\pi(W_0,\vec a,\vec b)\text{\,d}\vec a \text{\,d}\vec b
\\&= \frac{1}{(2\pi)^{k}}e^{\left[tr(W_0\log P)+tr\left\{(1-W_0)\log(J-P)\right\}\right]}\int_{\mathcal N_{\gamma^*}\{(\vec a_0,\vec b_0)\}}\!\!\exp\left\{-\frac{||\vec a||_2^2+||\vec b||_2^2}{2}\right\}  \text{\,d}\vec a \text{\,d}\vec b >0.
\end{aligned}
\label{equ: 3}	
\end{equation}
Combining (\ref{equ: 1}), (\ref{equ: 2}) and (\ref{equ: 3}) leads to the conclusion.
\end{proof}

\subsection*{Details of Sampling Algorithms}

In step (i), let the elements of $(\theta,\{\beta_{i,j}\}_{(i,j)})$ be represented by a vector $\vec s$. At the beginning, we simulate a velocity vector $\vec v$ (with equal length to $\vec s$) from $\No(\vec 0, \Omega^{-1})$ (with $\Omega$ positive definite matrix); using the value of $(\vec s,\vec v)$ as initial state $(\vec s^0,\vec v^0)$ at time $t=0$, we simulate the Hamiltonian dynamics defined by
\be
\frac{\partial s^t_i}{\partial t} = [\Omega v^t]_i,\qquad 
\frac{\partial v^t_i}{\partial t} = -\frac{\partial U}{\partial q^t_i}.
\ee
We use the leap-frog algorithm to obtain the approximate solution to the above equations, $(\vec s^h,\vec v^h)$ at the time $t=h$. When $\vec s^h$ satisfy the associated rank$(X)=k$, we treat them as proposal $(\vec s^*,\vec v^*)$, and use the Metropolis-Hastings criterion, if
\be
u<  \exp \{-U(\vec s^*)+U(\vec s)
-\vec v^{\ast\rm T}\Omega \vec v^\ast/2
+ \vec v\T\Omega \vec v/2 
\} ,\; u\sim \text{Uniform}(0,1),
\ee
we accept and set $\vec s:=\vec s^*$; otherwise, we reject the proposal and keep the original value.

In step (ii), we use the exchange algorithm. Specifically, we first propose a new set of $(\vec a ^*,\vec b ^*,\vec p^*)$ via
\be
&p^*_{j}\sim \text{Beta}(1+\sum_i w_{i,j},1+ n- \sum_i w_{i,j}), \\
& a^*_j\sim \text{Uniform} (a_j- r_j, a_j + r_j),
b^*_j\sim \text{Uniform} (b_j- r_j, b_j + r_j),
\ee
with $r_j>0$. Then we generate a new binary matrix $Y\in \{0,1\}^{n\times k}$ from density $g(Y; \vec p^*)/Z(\vec a^*,\vec b^*,\vec p^*)$ using rejection sampling: sample a matrix $Y^*$ with $y^*_{i,j}\sim \text{Bernoulli}(p^*_{j})$, and accept it as $Y$ only when rank$\{ Y^*\diag\{a^*_j\}+(J-Y^*)\diag\{b^*_j\}) \}=k$, otherwise repeat.  Considering the exchange of $(\vec a,\vec b,\vec p)$ and $(\vec a^*,\vec b^*,\vec p^*)$ between the posterior and $g(Y; \vec p^*)/Z(\vec a^*,\vec b^*,\vec p^*)$ , we accept the proposal $(\vec a, \vec b, \vec p):=(\vec a^*, \vec b^*, \vec p^*)$ for the iteration, if
\be
u<  \frac{
\exp\{ -U (\theta,W, \vec a^*, \vec b^*, \vec p^*)\}/{Z(\vec a^*, \vec b^*, \vec p^*)}} 
{
\exp\{ -U (\theta,W, \vec a, \vec b, \vec p)\}/{Z(\vec a, \vec b, \vec p)} 
)
}
\cdot
\frac{
g(W;\vec p) 
}
{
g(W;\vec p^*)
}  
\cdot
\frac{
g(Y;\vec p)
/Z(\vec a, \vec b, \vec p)
}
{
g(Y;\vec p^*)
/Z(\vec a^*, \vec b^*, \vec p^*)
} ,
\quad
u\sim \text{Uniform}(0,1);
\ee
otherwise, we keep the original value. Note that all the $Z$'s are canceled in the above ratio. We provide the code on the above algorithm (including the adaptation of leap-frog stepsize, $\Omega$, $h$ and $r_j$).

The source code for the sampling algorithms with the application on brain network data is provided and maintained on \url{https://github.com/moran-xu/MultiscaleFactorHMC}.

\bibliography{reference} 

\begin{thebibliography}{}

\bibitem[\protect\citeauthoryear{Breiman}{Breiman}{2017}]{breiman2017classification}
Breiman, L. (2017).
\newblock {\em {Classification and Regression Trees}}.
\newblock Routledge.

\bibitem[\protect\citeauthoryear{Chikuse}{Chikuse}{2012}]{chikuse2012statistics}
Chikuse, Y. (2012).
\newblock {\em {Statistics on Special Manifolds}}, Volume 174.
\newblock Springer Science \& Business Media.

\bibitem[\protect\citeauthoryear{Chipman, George, and McCulloch}{Chipman
  et~al.}{1998}]{chipman1998bayesian}
Chipman, H.~A., E.~I. George, and R.~E. McCulloch (1998).
\newblock {Bayesian CART Model Search}.
\newblock {\em Journal of the American Statistical Association\/}~{\em
  93\/}(443), 935--948.

\bibitem[\protect\citeauthoryear{Choi and Baraniuk}{Choi and
  Baraniuk}{2001}]{choi2001multiscale}
Choi, H. and R.~G. Baraniuk (2001).
\newblock {Multiscale Image Segmentation using Wavelet-Domain Hidden Markov
  Models}.
\newblock {\em IEEE Transactions on Image Processing\/}~{\em 10\/}(9),
  1309--1321.

\bibitem[\protect\citeauthoryear{Fosdick and Hoff}{Fosdick and
  Hoff}{2014}]{fosdick2014separable}
Fosdick, B.~K. and P.~D. Hoff (2014).
\newblock {Separable Factor Analysis with Applications to Mortality Data}.
\newblock {\em Annals of Applied Statistics\/}~{\em 8\/}(1), 120.

\bibitem[\protect\citeauthoryear{Ghosh and Ramamoorthi}{Ghosh and
  Ramamoorthi}{2003}]{ghosh2003bayesian}
Ghosh, J.~K. and R.~Ramamoorthi (2003).
\newblock {\em {Bayesian nonparametrics}}.
\newblock Springer Science \& Business Media.

\bibitem[\protect\citeauthoryear{Guhaniyogi, Qamar, and Dunson}{Guhaniyogi
  et~al.}{2017}]{guhaniyogi2017bayesian}
Guhaniyogi, R., S.~Qamar, and D.~B. Dunson (2017).
\newblock {Bayesian Tensor Regression}.
\newblock {\em Journal of Machine Learning Research\/}~{\em 18\/}(1),
  2733--2763.

\bibitem[\protect\citeauthoryear{Hoff}{Hoff}{2016}]{hoff2016equivariant}
Hoff, P.~D. (2016).
\newblock {Equivariant and Scale-free Tucker Decomposition Models}.
\newblock {\em Bayesian Analysis\/}~{\em 11\/}(3), 627--648.

\bibitem[\protect\citeauthoryear{Ma}{Ma}{2017}]{ma2017adaptive}
Ma, L. (2017).
\newblock {Adaptive Shrinkage in P{\'o}lya Tree Type Models}.
\newblock {\em Bayesian Analysis\/}~{\em 12\/}(3), 779--805.

\bibitem[\protect\citeauthoryear{Maddison, Mnih, and Teh}{Maddison
  et~al.}{2017}]{maddison2016concrete}
Maddison, C.~J., A.~Mnih, and Y.~W. Teh (2017).
\newblock {The Concrete Distribution: A Continuous Relaxation of Discrete
  Random Variables}.
\newblock {\em International Conference on Learning Representations\/}.

\bibitem[\protect\citeauthoryear{Murray, Ghahramani, and MacKay}{Murray
  et~al.}{2006}]{murrayexchange}
Murray, I., Z.~Ghahramani, and D.~J.~C. MacKay (2006).
\newblock {MCMC for Doubly-Intractable Distributions}.
\newblock In {\em Proceedings of the Twenty-Second Conference on Uncertainty in
  Artificial Intelligence}, UAI’06, Arlington, Virginia, USA, pp.\
  359–366. AUAI Press.

\bibitem[\protect\citeauthoryear{Nishimura, Dunson, and Lu}{Nishimura
  et~al.}{2020}]{nishimura2017discontinuous}
Nishimura, A., D.~Dunson, and J.~Lu (2020).
\newblock {Discontinuous Hamiltonian Monte Carlo for Sampling Discrete
  Parameters}.
\newblock {\em Biometrika\/}~{\em (in press)}.

\bibitem[\protect\citeauthoryear{Norouzi, Collins, Johnson, Fleet, and
  Kohli}{Norouzi et~al.}{2015}]{norouzi2015efficient}
Norouzi, M., M.~Collins, M.~A. Johnson, D.~J. Fleet, and P.~Kohli (2015).
\newblock {Efficient Non-Greedy Optimization of Decision Trees}.
\newblock In {\em Advances in Neural Information Processing Systems}, pp.\
  1729--1737.

\bibitem[\protect\citeauthoryear{Salter-Townshend and
  McCormick}{Salter-Townshend and McCormick}{2017}]{salter2017latent}
Salter-Townshend, M. and T.~H. McCormick (2017).
\newblock {Latent Space Models for Multiview Network Data}.
\newblock {\em The annals of applied statistics\/}~{\em 11\/}(3), 1217.

\bibitem[\protect\citeauthoryear{Wong, Ma, et~al.}{Wong
  et~al.}{2010}]{wong2010optional}
Wong, W.~H., L.~Ma, et~al. (2010).
\newblock {Optional P{\'o}lya tree and Bayesian inference}.
\newblock {\em The Annals of Statistics\/}~{\em 38\/}(3), 1433--1459.

\bibitem[\protect\citeauthoryear{Wu, Tjelmeland, and West}{Wu
  et~al.}{2007}]{wu2007bayesian}
Wu, Y., H.~Tjelmeland, and M.~West (2007).
\newblock {Bayesian CART: Prior Specification and Posterior Simulation}.
\newblock {\em Journal of Computational and Graphical Statistics\/}~{\em
  16\/}(1), 44--66.

\end{thebibliography}
\bibliographystyle{plain}
 
\end{document}